\documentclass[preprint,number]{elsarticle}
\usepackage{epsfig}
\usepackage{amsmath,graphicx}
\usepackage{mathrsfs}
\usepackage{amsbsy}
\usepackage{amssymb}
\usepackage{amsfonts}
\usepackage[english]{babel}
\usepackage{fancyhdr}
\begin{document}
\title{Spin motive forces and current fluctuations due to Brownian motion of domain walls}
\author{M.E. Lucassen\corref{cor1}}
\ead{m.e.lucassen@uu.nl}
\author{R.A. Duine}
\address{Institute for Theoretical Physics, Utrecht University,
Leuvenlaan 4, 3584 CE Utrecht, The Netherlands}
\date{\today}
\begin{abstract}
We compute the power spectrum of the noise in the current due to
spin motive forces by a fluctuating domain wall. We find that the
power spectrum of the noise in the current is colored, and depends
on the Gilbert damping, the spin transfer torque parameter
$\beta$, and the domain-wall pinning potential and magnetic
anisotropy. We also determine the average current induced by the
thermally-assisted motion of a domain wall that is driven by an
external magnetic field. Our results suggest that measuring the
power spectrum of the noise in the current in the presence of a
domain wall may provide a new method for characterizing the
current-to-domain-wall coupling in the system.
Semiconductors; D. Noise\\
\\
Pacs numbers: 72.15 Gd, 72.25 Pn, 72.70 +m
\end{abstract}
\maketitle
\section{Introduction}
Voltage noise has long been considered a problem. Engineers have
been concerned with bringing down noise in electric circuits for
more than a century. The seminal work by Johnson\cite{johnson1928}
and Nyquist\cite{nyquist1928} on noise caused by thermal agitation
of electric charge carriers (nowadays called Johnson-Nyquist
noise) was largely inspired by the problem caused by noise in
telephone wires. The experimental work by Johnson tested the
earlier observations by engineers that noise increases with
increasing resistance in the circuit and increasing temperature.
He was able to show that there would always be a minimal amount of
noise, beyond which reduction of the noise is not possible, thus
providing a very practical tool for people working in the field.
At the same time, the theoretical support for these predictions
was given by Nyquist. It is probably not a coincidence that, at
the time of his research, Nyquist worked for the American
Telephone and Telegraph Company.

As long as noise is frequency-independent, i.e., white like
Johnson-Nyquist noise, it is indeed often little more than a
nuisance (a notable exception to this is shot
noise\cite{beenakker1997} at large bias voltage). However,
frequency-dependent, i.e., colored noise can contain interesting
information on the system at hand. For example, in a recent paper
Xiao {\it et al.}\cite{xiao2009} show that, via the mechanism of
spin pumping\cite{tserkovnyak2002}, a thermally agitated spin
valve emits noisy currents with a colored power spectrum. They
show that the peaks in the spectrum coincide with the precession
frequency of the free ferromagnet of the spin valve. This opens up
the possibility of an alternative measurement of the ferromagnetic
resonance frequencies and damping, where one does not need to
excite the system, but only needs to measure the voltage noise
power spectrum. Here, we see that properties of the noise contain
information on the system. Clearly, this proposal only works if
the Johnson-Nyquist noise is not too large compared to the colored
noise.

Not only precessing magnets in layered structures induce currents:
Recent theoretical work has increased interest in the inverse
effect of current-driven domain-wall motion, whereby a moving
domain wall induces an electric
current\cite{barnes2007,saslow2007,duine2008a,tserkovnyak2008}.
Experimentally, this effect has been seen recently with
field-driven domain walls in permalloy wires\cite{yang2009}. These
so-called spin motive forces ultimately arise from the same
mechanism as spin pumping induced by the precessing magnet in a
spin valve, i.e., both involve dynamic magnetization that induces
spin currents that are subsequently converted into a charge
current.

In this paper, we study the currents induced by domain walls at
nonzero temperature. In particular, we determine the (colored)
power spectrum of the emitted currents due to a fluctuating domain
wall, both in the case of an unpinned domain wall
(Sec.~\ref{subsec:unpinned}), and in the case of a domain wall
that is extrinsically pinned (Sec.~\ref{subsec:pinned}). We also
compute the average current induced by a field-driven domain wall
at nonzero temperature. We end in Sec.~\ref{sec:conclusion} with a
short discussion and, in particular, compare the magnitude of the
colored noise obtained by us with the magnitude of the
Johnson-Nyquist noise.

\section{Spin motive forces due to fluctuating domain
walls}\label{sec:powerspectrum} In this section, we compute the
power spectrum of current fluctuations due to spin motive forces
that arise when a domain wall is thermally fluctuating. We
consider separately the case of intrinsic and extrinsic pinning.

\subsection{Model and approach}\label{subsec:model}
\begin{figure}
\centering
\includegraphics[width=8.5cm]{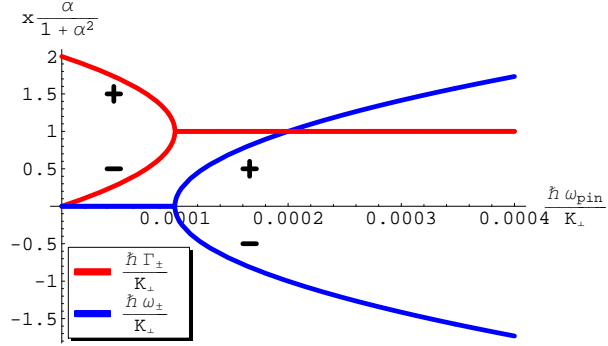}
\caption{Values of $\Gamma_\pm$ (red curves) and $\omega_\pm$
(blue curves) as a function of the pinning for $\alpha=0.02$
.}\label{fig:omegaengamma}
\end{figure}

The equations of motion for the position $X$ and the chirality
$\phi$ of a rigid domain wall at nonzero temperature are given
by\cite{tatara2004,duine2007b,lucassen2009}
\begin{align}\label{eq:EOM1}
\frac{\dot{X}}{\lambda}=&\alpha\dot{\phi}+\frac{K_\perp}{\hbar}\sin
2\phi +\sqrt{\frac{D}{2}}\eta_1\;,\\\label{eq:EOM2}
\dot{\phi}=&-\alpha\frac{\dot{X}}{\lambda} +F_{\rm pin
}+\sqrt{\frac{D}{2}}\eta_2\;,
\end{align}
where $\alpha$ is Gilbert damping, $K_\perp$ is the hard-axis
anisotropy, and $\lambda=\sqrt{K/J}$ is the domain-wall width,
with $J$ the spin stiffness and $K$ the easy-axis anisotropy. We
introduce a pinning force, denoted by $F_{\rm pin}$, to account
for irregularities in the material. We have assumed that the
pinning potential only depends on the position of the domain wall.
Pinning sites turn out to be well-described by a potential that is
quadratic in $X$, such that we can take $F_{\rm pin}=-2\omega_{\rm
pin}X/\lambda$\cite{tatara2004}. The Gaussian stochastic forces
$\eta_i$ describe thermal fluctuations and are determined by
\begin{align}\label{eq:thermalcorrelations}
\langle\eta_i(t)\rangle=0\;;
\qquad\langle\eta_i(t)\eta_j(t')\rangle=\delta_{ij}\delta(t-t')\;.
\end{align}
They obey the fluctuation-dissipation theorem\cite{duine2007b}
\begin{align}\label{eq:thermalstrength}
D=\frac{2\alpha k_{\rm B}T}{\hbar N_{\rm DW}}\;.
\end{align}
Note that in this expression, the temperature $T$ is effectively
reduced by the number of magnetic moments in the domain wall
$N_{\rm DW}=2\lambda A/a^3$, with $A$ the cross-sectional area of
the sample, and $a$ the lattice spacing. Up to linear order in the
coordinate $\phi$, valid when $K_\perp>k_{\rm B}T$, we can write
the equations of motion in
Eqs.~(\ref{eq:EOM1})~and~(\ref{eq:EOM2}) as
\begin{align}\label{eq:EOM3}
\partial_t\vec{x}=M\vec{x}+N\vec{\eta}\;,
\end{align}
where
\begin{align}
M=\frac{2}{1+\alpha^2}\begin{pmatrix}-\alpha\omega_{\rm
pin}&\frac{K_\perp}{\hbar}\\\\-\omega_{\rm pin}&-\alpha
\frac{K_\perp}{\hbar}\end{pmatrix}\;&;\quad
\vec{x}=\begin{pmatrix}\frac{X}{\lambda}\\\phi\end{pmatrix}\;,
\end{align}
and
\begin{align} N=\frac{1}{1+\alpha^2}\sqrt{\frac{\alpha k_{\rm B}T}{N_{\rm DW}\hbar}}
\begin{pmatrix}1&\alpha\\-\alpha&1\end{pmatrix}\;&;\quad
\vec{\eta}=\begin{pmatrix}\eta_1\\\eta_2\end{pmatrix}\;.
\end{align}
We readily find that the eigenfrequencies of the system,
determined by the eigenvalues $\Lambda_\pm$ of the matrix $M$, are
\begin{align}\label{eq:eigenfrequencies}
\Lambda_\pm\equiv&
\;i\omega_\pm-\Gamma_\mp=-\frac{\alpha}{1+\alpha^2}\left(\omega_{\rm
pin}+\frac{K_\perp}{\hbar}\right)\nonumber\\&\pm\frac{\alpha}{1+\alpha^2}
\sqrt{\left(\omega_{\rm pin
}-\frac{K_\perp}{\hbar}\right)^2-\frac{4}{\alpha^2}\omega_{\rm
pin}\frac{K_\perp}{\hbar}}\;,
\end{align}
with both the eigenfrequencies $\omega_\pm$ and their damping
rates $\Gamma_\pm$ real numbers. Their behavior as a function of
$\hbar\omega_{\rm pin}/K_\perp$ is shown in
Fig.~\ref{fig:omegaengamma}.
Note that this expression has an imaginary part for pinning
potentials that obey $\hbar\omega_{\rm
pin}/K_\perp\geq(\alpha/2)^2$, and, because for typical materials
the damping assumes values $\alpha\sim 0.01-0.1$, the
eigenfrequency assumes nonzero values already for very small
pinning potentials. Without pinning potential ($\omega_{\rm
pin}=0$) the eigenvalues are purely real-valued and the motion of
the domain wall is overdamped since $\Gamma_\pm\geq 0$ and
$\omega_\pm=0$.

If we include temperature, we find from the solution of
Eq.~(\ref{eq:EOM3}) [without loss of generality we choose
$X(t=0)=\phi(t=0)=0$] that the time derivatives of the collective
coordinates are given by
\begin{align}
\partial_t\vec{x}(t)=M e^{Mt}\int_0^tdt'e^{-Mt'}N\vec{\eta}(t')
+N\vec{\eta}(t)\;,
\end{align}
for one realization of the noise. By averaging this solution over
realizations of the noise, we compute the power spectrum of the
current induced by the domain wall under the influence of thermal
fluctuations as follows.

It was shown by one of us\cite{duine2008a} that up to linear order
in time derivatives, the current induced by a moving domain wall
is given by
\begin{align}\label{eq:current}
I(t)=-\frac{A\hbar}{|e|L}(\sigma_\uparrow-\sigma_\downarrow)
\left[\dot{\phi}(t)-\beta\frac{\dot{X}(t)}{\lambda}\right]\;,
\end{align}
with $L$ the length of the sample, and $\beta$ the sum of the
phenomenological dissipative spin transfer torque
parameter\cite{zhang2004} and non-adiabatic contributions. The
power spectrum is defined as
\begin{align}
P(\omega)=2\int_{-\infty}^{+\infty}d(t-t')
e^{-i\omega(t-t')}\langle I(t)I(t')\rangle\;.
\end{align}
Note that in this definition the power spectrum has units
$[P]={\rm A^2/Hz}$, not to be mistaken with the power spectrum of
a voltage-voltage correlation, which has units $[P]={\rm V^2/Hz}$.
In both cases, however, the power spectrum can be seen as a
measure of the energy output per frequency interval. We introduce
now the matrix
\begin{align}
O=\left(\frac{A\hbar}{|e|L}\right)^{2}
(\sigma_\uparrow-\sigma_\downarrow)^{2}
\begin{pmatrix}\beta^2&-\beta\\-\beta&1\end{pmatrix}\;,
\end{align}
so that we can write the correlations of the current as
\begin{align}\label{eq:currentcorrelations}
&\langle I(t)I(t')\rangle= \Big\langle[\partial_t\vec{x}(t)]^{\rm
T} O\partial_{t'}\vec{x}(t')\Big\rangle=\nonumber\\
&\int_0^t\int_0^{t'}dt''dt'''\Big\langle\vec{\eta}(t'')^{\rm
T}N^{\rm T}e^{M^{\rm T}(t-t'')}M^{\rm T}OMe^{M(t'-t''')}N
\vec{\eta}(t''')\Big\rangle\nonumber\\
&+\int_0^{t}dt''\Big\langle\vec{\eta}(t'')^{\rm T}N^{\rm
T}e^{M^{\rm T}(t-t'')}M^{\rm T}ON\vec{\eta}(t')\Big\rangle
\nonumber\\
&+\int_0^{t'}dt''\Big\langle \vec{\eta}(t)^{\rm T}N^{\rm T
}OMe^{M(t'-t'')}N\vec{\eta}(t'')\Big\rangle +\Big\langle
\vec{\eta}(t)^{\rm T}N^{\rm
T}ON\vec{\eta}(t')\Big\rangle=\nonumber\\
&\theta(t-t')\Bigg\{\int_0^{t'}dt''{\rm Tr}\Big[N^{\rm T}e^{M^{\rm
T}(t-t'')}M^{\rm T}OMe^{M(t'-t'')}N\Big]\nonumber\\
&\qquad\qquad\qquad+{\rm Tr}\Big[N^{\rm T}e^{M^{\rm
T}(t-t')}M^{\rm
T}ON\Big]\Bigg\}\nonumber\\
&+\theta(t'-t)\Bigg\{\int_0^{t}dt''{\rm Tr}\Big[N^{\rm T}e^{M^{\rm
T}(t-t'')}M^{\rm T}OMe^{M(t'-t'')}N\Big]\nonumber\\
&\qquad\qquad\qquad+{\rm Tr}\Big[N^{\rm T
}OMe^{M(t'-t)}N\Big]\Bigg\}+\delta(t-t'){\rm Tr}\Big[N^{\rm
T}ON\Big]\;.
\end{align}
We evaluate the traces that appear in this expression to find that
the power spectrum is given by
\begin{align}\label{eq:powerspectrum}
&P(\omega)=\nonumber\\
&2\left(\frac{A\hbar}{|e|L}\right)^{2}
\frac{(\sigma_\uparrow-\sigma_\downarrow)^{2}}{1+\alpha^2}
\frac{\alpha k_{\rm B}T}{\hbar N_{\rm DW}
}\times\Bigg[(1+\beta)^2-\Bigg\{(1+\beta^2)(1+\alpha^2)^2\Bigg(\frac{\hbar\omega_{\rm
pin}}{K_\perp}\Bigg)^2\nonumber\\
&-\Bigg[\beta^2-\alpha^2+2(1+\beta^2)\frac{\hbar\omega_{\rm
pin}}{K_\perp}+(1-\alpha^2\beta^2)\Bigg(\frac{\hbar\omega_{\rm
pin}}{K_\perp}\Bigg)^2\Bigg]\Bigg(\frac{\hbar\omega}{K_\perp}\frac{1+\alpha^2}
{2}\Bigg)^2\Bigg\}\Big{/}\nonumber\\
&\Bigg\{(1+\alpha^2)^2\Bigg(\frac{\hbar\omega_{\rm
pin}}{K_\perp}\Bigg)^2 +\Bigg[\alpha^2-2\frac{\hbar\omega_{\rm
pin}}{K_\perp}+\alpha^2\Big(\frac{\hbar\omega_{\rm pin
}}{K_\perp}\Bigg)^2\Bigg]\Bigg(\frac{\hbar\omega}{K_\perp}\frac{1+\alpha^2}
{2}\Bigg)^2\nonumber\\
& +\Bigg(\frac{\hbar\omega}{K_\perp}\frac{1+\alpha^2}
{2}\Bigg)^4\Bigg\}\Bigg]\;.
\end{align}

\subsection{Domain wall without extrinsic pinning}\label{subsec:unpinned}
\begin{figure}[h!]
\centering
\includegraphics[width=9cm]{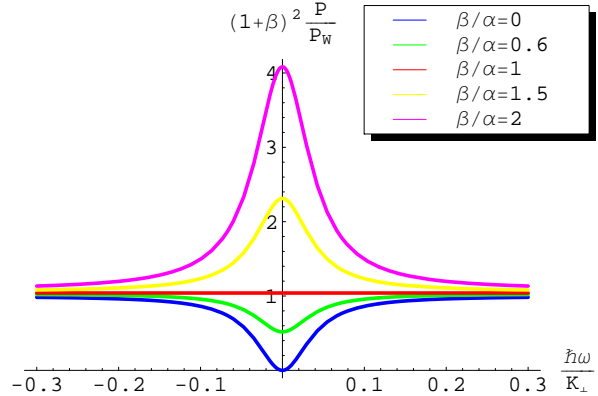}
\caption{The power spectrum for $\alpha=0.02$ and several values
of $\beta$.}\label{fig:powerspec}
\end{figure}

We first consider a domain wall with $F_{\rm pin}=0$. In this
case, only the chirality $\phi$ determines the energy, a situation
referred to as intrinsic pinning\cite{tatara2004}. From the result
in Eq.~(\ref{eq:powerspectrum}) we find that the power spectrum is
given by
\begin{align}\label{eq:powerspecs}
&P(\omega)=2\left(\frac{A\hbar}{|e|L}\right)^{2}
\frac{(\sigma_\uparrow-\sigma_\downarrow)^{2}}{1+\alpha^2}
\frac{\alpha k_{\rm
B}T}{\hbar N_{\rm DW}}\times\nonumber\\
&\Bigg\{(1+\beta)^2 +\frac{\beta^2-\alpha^2}{\alpha^2}
\Big[1+\Big(\frac{\hbar\omega}{K_\perp}\frac{1+\alpha^2}
{2\alpha}\Big)^2\Big]^{-1}\Bigg\}\;.
\end{align}
Indeed, we find that next to a constant contribution there is also
a frequency-dependent contribution for $\beta\neq\alpha$, i.e.,
the power spectrum is colored. The fact that $\beta=\alpha$ is a
special case is understood from the fact that in that case we have
macroscopic Galilean invariance. This translates to white noise in
the current correlations. The power spectrum is a Lorentzian,
centered around $\omega=0$ because the domain wall is overdamped
in this case, with a width determined by the damping rate in
Eq.~(\ref{eq:eigenfrequencies}) as
$\hbar\Gamma_+/K_\perp=2\alpha/(1+\alpha^2)$. Relative to the
white-noise contribution
\begin{align}
P_{\rm W}=2(1+\beta)^2\left(\frac{A\hbar}{|e|L}\right)^{2}
\frac{(\sigma_\uparrow-\sigma_\downarrow)^{2}}{1+\alpha^2}
\frac{\alpha k_{\rm B}T}{\hbar N_{\rm DW}}\;,
\end{align}
the height of the peak is given by $\Delta P=P_{\rm W
}(\beta^2-\alpha^2)/\alpha^2$. The behavior of the power spectrum
is illustrated in Fig.~\ref{fig:powerspec} for several values of
$\beta/\alpha$.

\subsection{Extrinsically pinned domain wall}\label{subsec:pinned}

\begin{figure}[htp]
\centering
\begin{tabular}{c}
\includegraphics[width=9cm]{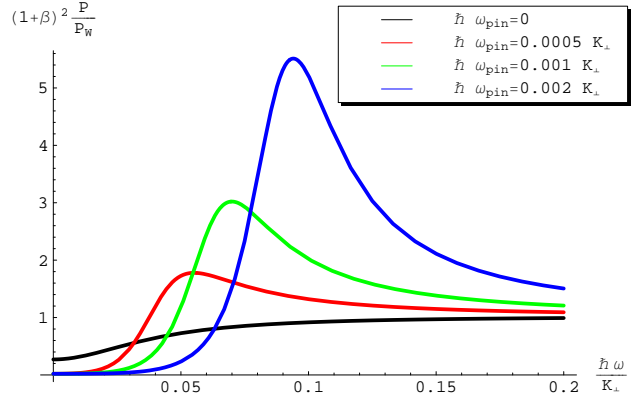}\\(a)\\
\includegraphics[width=9cm]{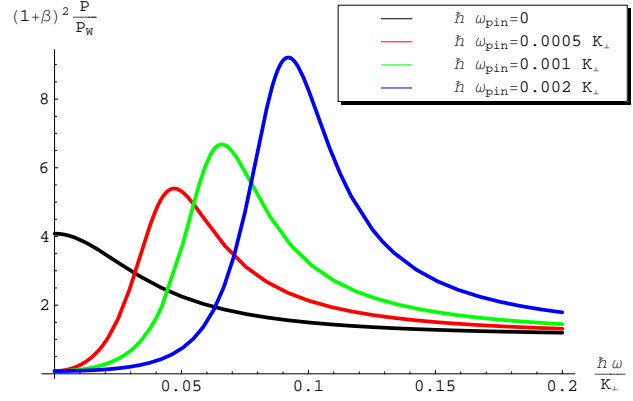}\\(b)\\
\includegraphics[width=9cm]{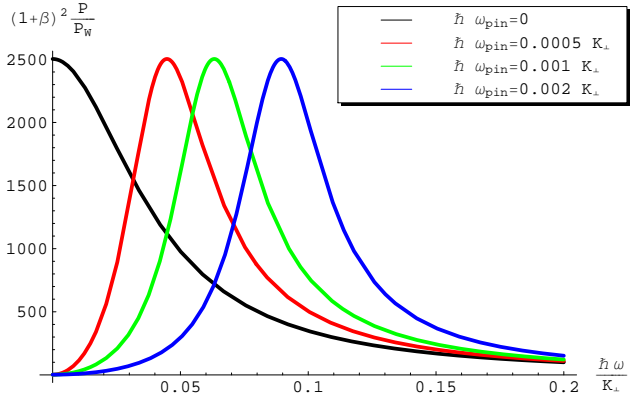}\\(c)
\end{tabular}
\caption{The power spectrum as a function of the frequency
$\omega$ and the pinning potential $\omega_{\rm pin}$ for
$\beta=\alpha/2$ (a), $\beta=2\alpha$ (b) and $\beta=50\alpha$
(c), all for $\alpha=0.02$.}\label{fig:powerspectrum}
\end{figure}

For extrinsically pinned domain walls the behavior of the power
spectrum given by Eq.~(\ref{eq:powerspectrum}) is depicted in
Fig.~\ref{fig:powerspectrum}.
We see that for $\hbar\omega_{\rm pin}/K_\perp\gtrsim \alpha^{2}$
the peaks in the power spectrum are approximately centered around
the eigenfrequencies
$\hbar\omega/K_\perp\simeq\pm2\sqrt{\hbar\omega_{\rm
pin}/K_\perp}$, consistent with Eq.~(\ref{eq:eigenfrequencies}).
We can discern between two regimes, one where $\beta\sim\alpha$ in
figs.~\ref{fig:powerspectrum} (a-d), and one where
$\beta\gg\alpha$ in figs.~\ref{fig:powerspectrum} (e-f). In the
former regime, the height of the peaks in the power spectrum
depend strongly on the pinning. For small $\omega_{\rm pin}$ we
see a clear dependence on the value of $\beta$, whereas for large
$\omega_{\rm pin}$ this dependence is less significant. In the
regime of large $\beta$, the height of the peaks hardly depends on
the pinning and is approximately given by $P\simeq P_{\rm
W}\beta^2/\alpha^2$. Note that the width of the peaks is
independent of $\beta$. For pinning potentials
$\alpha^2<\hbar\omega_{\rm pin}/K_\perp$ the width is given by
$\hbar\Gamma/K_\perp=\alpha(1+\hbar\omega_{\rm
pin}/K_\perp)/(1+\alpha^2)$, so for $\hbar\omega_{\rm pin}\ll
K_\perp$ the dependence of the width on the pinning is negligible.

\section{Spin motive forces due to thermally-assisted field-driven domain walls}\label{sec:inducedcurrent}
\begin{figure}[h!]
\centering
\includegraphics[width=9cm]{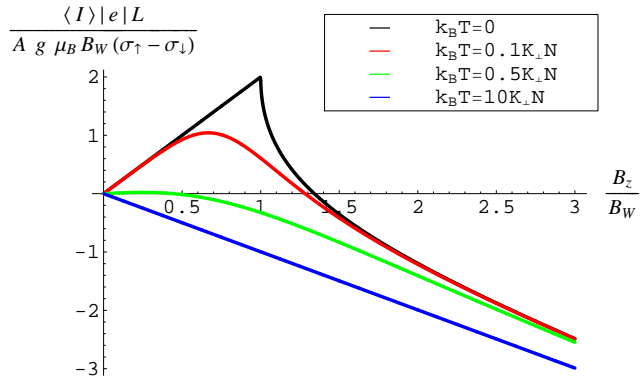}
\caption{The average current induced by a domain wall as a
function of the magnetic field applied to this domain wall. We
show curves for several temperatures, with $\alpha=0.02$ and
$\beta=2\alpha$. The normalization of the axes $B_{\rm W}$ is the
Walker-breakdown field which is given by $B_{\rm W}=\alpha
K_{\perp}/g\mu_{\rm B}$. }\label{fig:averagecurrent}
\end{figure}

In this section we compute the average current that is induced by
a moving domain wall. The domain wall is moved by applying an
external magnetic field parallel to the easy axis of the sample.
In this section we take into account temperature but ignore
extrinsic pinning (see Ref.~\cite{lecomte2009} for a calculation
of spin motive forces in a weakly extrinsically pinned system for
$\beta=0$). The equations of motion for a field-driven domain wall
are then given by\cite{tatara2004,duine2007b,lucassen2009}
\begin{align}\label{eq:EOM4}
\frac{\dot{X}}{\lambda}&=\alpha\dot{\phi}+\frac{K_\perp}{\hbar}\sin
2\phi +\sqrt{\frac{D}{2}}\eta_1\;,\\\label{eq:EOM5}
\dot{\phi}&=-\alpha\frac{\dot{X}}{\lambda} +\frac{g\mu_{\rm
B}B_z}{\hbar} +\sqrt{\frac{D}{2}}\eta_2\;,
\end{align}
where $gB_z$ is the magnetic field in the z direction, and the
stochastic forces are determined by
Eq.~(\ref{eq:thermalcorrelations}) with the strength given by
Eq.~(\ref{eq:thermalstrength}). In earlier
work\cite{lucassen2009}, we computed from these coupled equations
the average drift velocity of a domain wall (here, we set the
applied spin current zero)
\begin{align}\label{eq:driftvelocity}
\alpha\frac{\langle\dot{X}\rangle}{\lambda}=
-\langle\dot{\phi}\rangle+\frac{g\mu_{\rm B}B_z}{\hbar}\;,
\end{align}
with (we omit factors $1+\alpha^2\simeq1$)
\begin{align}\label{eq:driftvelocityangle}
&\langle\dot{\phi}\rangle=-2\pi\frac{\alpha k_{\rm B}T}{\hbar
N_{\rm DW}}\big(e^{-2\pi\frac{ g\mu_{\rm B}B_zN_{\rm DW}}{\alpha
k_{\rm
B}T}}-1\big)\Big{/}\nonumber\\
&\Bigg\{\int_0^{2\pi} d \phi e^{\frac{N_{\rm DW}}{k_{\rm
B}T}\big(\frac{g\mu_{\rm B}B_z}{\alpha}\phi+\frac{K_\perp }{2}\cos
2\phi\big)}\Bigg[\int_0^{2\pi}d \phi' e^{-\frac{N_{\rm DW}}{k_{\rm
B}T}\big(\frac{g\mu_{\rm
B}B_z}{\alpha}\phi'+\frac{K_\perp}{2}\cos 2\phi'\big)}\nonumber\\
&\qquad\qquad+\big(e^{-2\pi \frac{g\mu_{\rm B}B_z N_{\rm
DW}}{\alpha k_{\rm B}T}}-1\big)\int_{0}^{\phi}
d\phi'e^{-\frac{N_{\rm DW}}{k_{\rm B}T}\big(\frac{g\mu_{\rm
B}B_z}{\alpha}\phi' +\frac{K_\perp}{2}\cos
2\phi'\big)}\Bigg]\Bigg\}\;.
\end{align}
Combining Eqs.~(\ref{eq:current})~and~(\ref{eq:driftvelocity})
gives us the average current straightforwardly
\begin{align}
\langle I\rangle
=-\frac{A\hbar}{\alpha|e|L}(\sigma_\uparrow-\sigma_\downarrow)
\left[\left(\alpha+\beta\right)\langle\dot{\phi}\rangle
-\beta\frac{g\mu_{\rm B}B_z}{\hbar}\right]\;.
\end{align}
We evaluate this expression for several temperatures in
Fig.~\ref{fig:averagecurrent}.
The black curve in Fig.~\ref{fig:averagecurrent} is computed at
zero temperature. It increases linearly with the field up to the
Walker-breakdown field $B_{\rm W}=\alpha K_\perp/g\mu_{\rm B}$,
where it reaches a maximum current of $\langle
I\rangle|e|L/Ag\mu_{\rm B}B_{\rm
W}(\sigma_\uparrow-\sigma_\downarrow)=\beta/\alpha$. Then, it
drops and even changes sign. This curve is consistent with the
curves obtained by one of us\cite{duine2008a}. There, an open
circuit is treated where there is no current but a voltage. The
voltage is related to the current as $\Delta V=\langle I\rangle
L/A(\sigma_\uparrow+\sigma_\downarrow)$, such that indeed $\Delta
V/V_0=\langle I\rangle|e|L/Ag\mu_{\rm B}B_{\rm
W}(\sigma_\uparrow-\sigma_\downarrow)$, where the normalization is
defined as $V_0={\mathcal P} g \mu_{\rm B} B_{\rm W}/|e|$ and the
polarization is given by ${\mathcal
P}\equiv(\sigma_\uparrow-\sigma_\downarrow)
/(\sigma_\uparrow+\sigma_\downarrow)$. Increasing temperatures
smoothen the peak around the Walker-breakdown field, and for high
temperatures, the peak vanishes altogether. Therefore, for fields
smaller than the Walker-breakdown field and for small
temperatures, the thermal fluctuations tend to decrease the
average induced current. However, for higher temperatures, the
current reverses and increases again. For fields sufficiently
larger than the Walker-breakdown field, we always find a slight
increase of the current as a function of temperature.

\section{Discussion and conclusions}\label{sec:conclusion}

In Sec.~\ref{sec:powerspectrum} we computed power spectra for
domain walls, both with and without extrinsic pinning. In
ferromagnetic metals, the spin transfer torque parameter has
values $\beta\sim\alpha$, for which we see that for large pinning
$\hbar\omega_{\rm pin}/K_\perp\gtrsim\alpha^2$ the power spectra
only weakly depend on the spin transfer torque parameter $\beta$.
Therefore, determination of $\beta$ is only possible for very
small pinning potentials. In an ideally clean sample without
extrinsic pinning the height and sign of the peak in the power
spectrum can give a clear indication whether $\beta$ is smaller,
(approximately) equal or larger than $\alpha$.

\begin{figure}[h!]
\centering
\includegraphics[width=8.5cm]{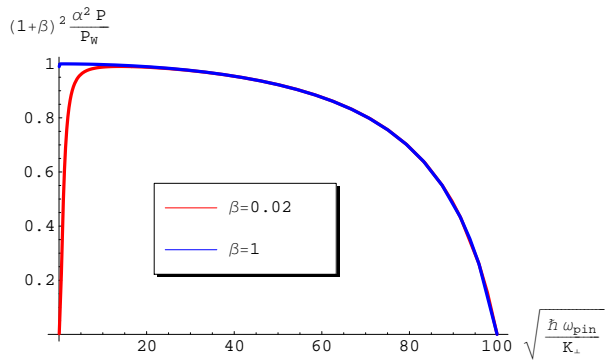}
\caption{The height of the peaks in the power spectrum as a
function of the pinning. We used $\alpha=0.02$ and two values for
$\beta$. We checked that the curves do not differ significantly
for different values for $\alpha$ and therefore, the maximal value
of the peaks in the power spectrum is $\Delta P(1+\beta^2)/P_{\rm
JN}\simeq \alpha^{-2}$. Note that the dependence on $\beta$ is
negligible for large pinning. This curve was obtained by inserting
the eigenfrequencies $\omega_\pm$ from
Eq.~(\ref{eq:eigenfrequencies}) in the expression for the power
spectrum in Eq.~(\ref{eq:powerspectrum}).}\label{fig:maximalpower}
\end{figure}

In order to perform these experiments, the contributions due to
the domain wall should not be completely overwhelmed by
Johnson-Nyquist noise. The ratio of the peaks in our power
spectrum and the Johnson-Nyquist noise determine the resolution of
the experiment, and we want it to be at least of the order of a
percent. We use that the resistance of the domain wall is small,
and the Johnson-Nyquist noise is governed by the resistance of the
wire. For a wire of length $L$ and cross-section $A$, the power
spectrum due to Johnson-Nyquist noise is given by $P_{\rm
JN}=4k_{\rm B}T(\sigma_\uparrow+\sigma_\downarrow)A/L$. From
section (\ref{subsec:unpinned}) we can estimate the ratio of the
height of the peak and the Johnson-Nyquist noise as $\Delta
P/P_{\rm JN}\simeq[(\beta/\alpha)^2-1]\alpha A\hbar
(\sigma_\uparrow+\sigma_\downarrow)/(4LN_{\rm DW}{\mathcal
P}^2e^2)$. To make a rough estimate of this ratio, we use
$\lambda\simeq 20 {\rm nm}$ such that $A/N_{\rm DW}\simeq 2\cdot
10^{-2}{\rm \AA}^2$ for $a\simeq 2{\rm \AA}$, a polarization
${\mathcal P}\simeq0.7$ and a conductivity
$\sigma_\uparrow+\sigma_\downarrow\simeq 10^7{\rm A/Vm}$. We then
find $\Delta P/P_{\rm JN}\simeq10^{-2}{\rm \AA}/L$, which shows
that the wire length must satisfy $L<1{\rm \AA}$ in order to have
a resolution of 1\%. We therefore conclude that this effect at
zero pinning is impossible to see in experiment, where wires are
typically at least of the order of $L\simeq 10 {\rm \mu m}$, i.e.,
five orders of magnitude larger. We can try to increase the signal
by turning on a pinning potential. Insertion of the
eigenfrequencies $\omega_\pm$ from Eq.~(\ref{eq:eigenfrequencies})
in Eq.~(\ref{eq:powerspectrum}) shows us that we can maximally
gain a factor $\alpha^{-2}\simeq 10^4$ for $\alpha=0.01$, as
illustrated in Fig.~\ref{fig:maximalpower}.
We can also still gain some resolution by considering a domain
wall in a nanoconstriction, where the width can be as small as
$\lambda=1$nm\cite{bruno1999}. All this adds up to a resolution of
$\Delta P/P_{\rm JN}\sim 1\%$. Therefore, with all parameter
values set to ideal values it appears to be possible to observe
our predictions in ferromagnetic metals, although the experimental
challenges are big.

In a magnetic semiconductor like GaMnAs  the number of magnetic
moments in the domain wall is two orders of magnitude smaller than
in a ferromagnetic conductor. This increases the ratio $\Delta
P/P_{\rm JN}$ with a factor $100$. However, the conductivity is
also considerably smaller than that in a ferromagnetic metal. The
conductivity of GaMnAs depends on many parameters, but a
reasonable estimate is that it is at least about three orders of
magnitude smaller than that of a metal, although usually even
smaller. Therefore, the advantage of a small number of magnetic
moments is cancelled. Another property, however, of GaMnAs is that
the parameter $\beta$ can assume considerably higher
values\cite{brataas2009}, up to $\beta=1$. This does not influence
the maximal value of the power spectrum, but it does dramatically
increase the value for small pinning.

Note that there are contributions to the noise that we did not
discuss in this article. For example, the time-dependent magnetic
field caused by a moving domain wall will induce electric currents
that contribute to the colored power spectrum. Distinguishing such
contributions from the spin motive forces was essential for the
experimental results by Yang {\it et al.} \cite{yang2009}, and
would also be important here.

In Sec.~\ref{sec:inducedcurrent}, we calculated the current
induced by a field-driven domain wall under the influence of
temperature. We estimate that in ferromagnetic metals at room
temperature that $k_{\rm B}T/K_\perp N\simeq10^{-3}$, which is
indistinguishable from the zero-temperature curve in
Fig.~\ref{fig:averagecurrent}. However, for magnetic
semiconductors, like GaMnAs, we find $k_{\rm B}T/K_\perp
N\simeq10^{-1}$ at $T=100{\rm K}$, which corresponds to the red
curve in Fig.~\ref{fig:averagecurrent}. We therefore expect
finite-temperature effects to be important in magnetic
semiconductors like GaMnAs.

\section{acknowledgement}
This work was supported by the Netherlands Organization for
Scientific Research (NWO) and by the European Research Council
(ERC) under the Seventh Framework Program. We would like to thank
Yaroslav Tserkovnyak for useful discussions.


\begin{thebibliography}{99}
\bibitem{johnson1928} J.B. Johnson, Nature {\bf 119}, 50 (1927); Phys. Rev. {\bf
29}, 367 (1927); Phys. Rev. {\bf 32}, 97 (1928).
\bibitem{nyquist1928} H. Nyquist, Phys. Rev. {\bf 29}, 614 (1927);
Phys. Rev. {\bf 32}, 110 (1928).
\bibitem{beenakker1997} M.J.M. de Jong and C.W.J. Beenakker, in
{\it Mesoscopic Electron Transport}, edited by L.L. Sohn, L.P.
Kouwenhoven and G. Schoen, NATO ASI Series (Kluwer Academic,
Dordrecht, 1997), Vol. 345, pp. 225-258.
\bibitem{xiao2009} J. Xiao, G.E.W. Bauer, S. Maekawa and A.
Brataas, Phys. Rev. B {\bf 79}, 174415 (2009).
\bibitem{tserkovnyak2002} Y. Tserkovnyak, A. Brataas and G.E.W.
Bauer, Phys. Rev. Lett. {\bf 88}, 117601 (2002).
\bibitem{barnes2007} S.E. Barnes and S. Maekawa, Phys. Rev. Lett.
{\bf 98}, 246601 (2007).
\bibitem{saslow2007} W.M. Saslow, Phys. Rev. B {\bf 76}, 184434
(2007).
\bibitem{duine2008a} R.A. Duine, Phys. Rev. B {\bf 77}, 014409
(2008); R.A. Duine, Phys. Rev. B {\bf 79}, 014407 (2009).
\bibitem{tserkovnyak2008} Y. Tserkovnyak and M. Mecklenburg, Phys.
Rev. B {\bf 77}, 134407 (2008).
\bibitem{yang2009} S.A. Yang, G.S.D. Beach, C. Knutson, D. Xiao,
Q. Niu, M. Tsoi, and J.L. Erskine, Phys. Rev. Lett. {\bf 102},
067201 (2009).
\bibitem{tatara2004} G. Tatara and H. Kohno, Phys. Rev. Lett. {\bf
92}, 086601 (2004); {\bf 96}, 189702 (2006).
\bibitem{duine2007b} R.A. Duine, A.S. N\'{u}\~{n}ez and A.H.
MacDonald, Phys. Rev. Lett. {\bf 98}, 056605 (2007).
\bibitem{lucassen2009} M.E. Lucassen, H.J. van Driel, C. Morais Smith and R.A. Duine,
Phys. Rev. B {\bf 79}, 224411 (2009).
\bibitem{zhang2004} S. Zhang and Z. Li, Phys. Rev. Lett. {\bf 93},
127204 (2004).
\bibitem{lecomte2009} V. Lecomte, S.E. Barnes, J.-P. Eckmann and
T. Giamarchi, Phys. Rev. B {\bf 80}, 054413 (2009).
\bibitem{bruno1999} P. Bruno, Phys. Rev. Lett. {\bf 83}, 2425
(1999).
\bibitem{brataas2009} K.M.D. Hals, A.K. Nguyen and A. Brataas, Phys. Rev. Lett. {\bf 102},
256601 (2009).
\end{thebibliography}
\end{document}